# Multimodal Wearable-Based Olfactory-Induced Emotion Recognition in Arousal-Valence Dimensions

Chen-Yang Xu[††], Lan Zhang[††], Fei-Yi Fan, Bin Hu, *Fellow IEEE*, Qing-Hao Meng*

***Abstract*—Olfaction is important for emotion regulation because it acts as a non-intrusive and cognitively lightweight pathway that directly engages the brain's affective circuitry and achieves unobtrusive emotional modulation. This trait is essential for advancing practical affective computing in daily and attention-critical scenarios. However, current olfactory emotion research has two key limitations. First, it overemphasises the valence dimension while neglecting arousal. Second, it lacks multimodal datasets that synchronously capture central and peripheral physiological responses to olfactory stimuli. To address these issues, we construct a large-scale multimodal olfactory emotion dataset based on 111 subjects, in which odors are labeled in the 2D arousal-valence space and electroencephalogram (EEG), electrocardiogram (ECG), and photoplethysmography (PPG) signals synchronously recorded. Nevertheless, multimodal signals present challenges such as non-stationarity, differences in latency, and cross-modal heterogeneity. Thus, we propose a spatiotemporal-frequency hybrid fusion network (STF-HFNet), which integrates three core modules. Frequency aggregation processing learns adaptive frequency aggregation in order to model non-stationary dynamics. Reciprocal guided attention enables reciprocal bidirectional calibration for cross-modal temporal alignment without synchronisation priors. Hybrid collaborative fusion combines spatial and channel attention mechanisms to enhance cross-modal complementarity while suppressing redundant information. Extensive experiments show that STF-HFNet achieves state-of-the-art (SOTA) recognition accuracies of 88.34% on the AMIGOS dataset and 92.40% on our self-constructed dataset, and outperform the SOTA methods by 8.27% and 5.07%, respectively.**

## I. Introduction

Wearable affective computing enables continuous, unobtrusive monitoring of emotional states in daily life and has broad potential for applications in areas such as stress management. In prior research, visual and auditory stimuli are frequently used to elicit emotions and the resulting changes in physiological signals are analysed. However, visual and auditory stimuli typically rely on the thalamocortical pathway for processing and require significant attentional resources, and thereby impose a considerable cognitive burden on users. This cognitive interference not only reduces the naturalness of emotional monitoring but may also pose additional safety risks in scenarios that demand high levels of attention, such as driving or surgery, as these situations require individuals to maintain full concentration. In contrast, the sense of smell is more closely linked to the brain regions responsible for emotional processing. Consequently, odors can influence emotions in a more natural and less intrusive manner, without readily disrupting the user's current focus.

Despite its unique advantages, olfactory affective computing still presents two critical bottlenecks. First, unlike visual stimuli [1] [2], odor stimuli [3] still lack standardised efforts for systematic mapping onto the 2-dimensional arousal–valence space, which makes it difficult to achieve unified representation and reliable evaluation of olfactory-induced emotions across both dimensions. Second, current olfactory datasets primarily collect electroencephalogram (EEG) signals [3], but often neglect peripheral physiological signals such as electrocardiogram (ECG) and photoplethysmography (PPG). Peripheral cardiovascular signals are crucial for describing a complete emotional response. Consequently, the lack of synchronised recordings of central and peripheral signals limits the comprehensive modelling of olfactory-induced emotions.

To address these issues, we constructed a multimodal olfactory emotion dataset that synchronises EEG, ECG, and PPG signals. However, the nature of multimodal physiological signals presents new methodological challenges because the three modalities are inherently heterogeneous. EEG contains high-dimensional and semantically rich information, whereas ECG and PPG provide low-dimensional but pattern-stable features. If the same modelling approach is applied to these modalities, it is difficult to fully exploit their respective strengths and complementary information. Furthermore, temporal inconsistencies exist between the different modalities. Due to physiological delays caused by neural conduction and haemodynamic processes, the responses of ECG and PPG typically lag behind those of the olfactory-evoked EEG [4]. Furthermore, olfactory-evoked multimodal physiological signals exhibit distinct non-stationary characteristics. These signals display time-varying frequency-domain dynamics and global frequency dependence—features that are often difficult to capture using local convolution or attention mechanisms.

While prior multimodal emotion recognition methods demonstrate progress, they often fail to explicitly model the coupled challenges inherent in multimodal physiological signals; these include non-stationary spectral dynamics,

* Corresponding author. [††] These authors contributed equally to this work.

This research is supported by the National Natural Science Foundation of China (Grant Nos 62203321 and 62571368).

Chen-Yang Xu, Lan Zhang, Qing-Hao Meng are with the School of Electrical and Information Engineering, Tianjin University, Tianjin, 300072, China (e-mail: xuchenyang@tju.edu.cn, zl2022@tju.edu.cn, qh_meng@tju.edu.cn).

Fei-Yi Fan is with Institute of Computing Technology, Chinese Academy of Sciences, Beijing, 100190, China. (e-mail: fanfeiyi@ict.ac.cn).

Bin Hu is with School of Medical Technology, Beijing Institute of Technology, Beijing 100081, China. (email: bh@bit.edu.cn).

physiological latency mismatch, and modality heterogeneity. These factors collectively undermine the reliability of cross-modal fusion for EEG, ECG, and PPG. Although techniques such as cross-modal transformers [5], hierarchical contrastive alignment [6], and multi-head cross-attention [7] yield substantial improvements, even state-of-the-art approaches such as uncertainty-aware graph contrastive networks exhibit several key deficiencies. Specifically, (1) frequency-domain information is frequently treated as static handcrafted input, which fails to characterise non-stationary spectral dynamics in an end-to-end manner. (2) Existing methods typically assume strict temporal alignment and overlook inherent physiological propagation delays across distinct modalities, a problem that is particularly acute in datasets lacking hardware synchronisation. And (3) uniform fusion strategies neglect intrinsic discrepancies in sampling rates as well as variations in structural morphology and detail granularity across modalities.

To address the aforementioned data scarcity and methodological gaps, we first construct the All-age Multimodal Olfactory Response & Emotion (AMORE) dataset. This is a synchronised multimodal (EEG-ECG-PPG) olfactory emotion dataset that fills the gap in existing olfactory multimodal resources by employing odor stimuli evaluated along the arousal-valence dimension. However, due to the heterogeneity and diversity of data, these issues must be addressed in order to enable effective multimodal data analysis. We therefore propose the spatiotemporal-frequency hybrid fusion network (STF-HFNet), whose architecture is explicitly co-designed to match the dataset's characteristics. (i) A frequency aggregation processing (FAP) module that performs fast Fourier transform (FFT)-based spectral reweighting with multi-head patch decomposition, which enables end-to-end modelling of non-stationary frequency dynamics rather than relying on static handcrafted frequency features. (ii) A reciprocal guided attention (RGA) mechanism that uses peripheral signal features as queries and keys to retrieve and reorganise EEG semantic representations, and explicitly compensates for physiological propagation delays instead of assuming temporal alignment. And (iii) a hybrid collaborative fusion (HCF) module that mitigates entanglement in multimodal correlations by integrating EEG channel attention with peripheral spatiotemporal attention and generates a joint gating mask to selectively enhance valuable features and suppress redundant features.

To summarise, the main contributions of this work are as follows:

- We designed an olfactory-induced paradigm based on the arousal-valence model and used it to screen for specific odors. We collected high-quality EEG, ECG and PPG signals from 111 subjects using wearable devices, which we used to construct the AMORE dataset.
- We proposed the STF-HFNet for multimodal emotion recognition, which has three properties. First, it stabilises the non-stationary frequency features of signals via frequency aggregation. Second, it compensates for central-peripheral propagation delays through peripheral-guided EEG recalibration. And third, it fuses heterogeneous modalities via a hybrid attention guidance map that selectively enhances complementary cues while suppressing modality-specific noise.
- We conducted extensive experiments on a public dataset and our self-collected dataset. The results demonstrate that STF-HFNet achieves recognition accuracies of 88.34% and 92.40%, respectively, and significantly outperform state-of-the-art (SOTA) methods. Ablation studies and visualisation analyses further verify the effectiveness of each module in the network.

## II. Related works

In this section, we present related work in two areas. (1) Olfactory-induced emotion recognition, which focuses on the variations in physiological signals triggered by olfactory stimuli, and (2) multimodal physiological signal-based emotion recognition methods under visual stimuli. Since multimodal research on olfactory-induced emotion recognition is limited, this paper reviews multimodal methods developed for visual stimuli.

### A. Olfactory-induced emotion recognition

Research in olfactory affective computing has shown that odor stimuli can effectively elicit measurable emotional responses through physiological signals, and particularly EEG signals. Xue et al. [8] validated the feasibility of decoding olfactory-induced emotions from EEG by developing an odor-video elicited physiological signal dataset that comprises EEG recordings from 10 subjects exposed to 32 video clips paired with 10 distinct odors. The brain topography analysis revealed enhanced orbitofrontal cortex activity under olfactory stimulation, which corroborates functional MRI findings on odor-emotion neural associations. Wu et al. [9] investigated the temporal dynamics information of olfactory-enhanced emotion induction and found that early-stage odor presentation significantly improved classification performance compared with video-only paradigms, particularly for negative emotions. Zhong et al. [10] extended this to virtual reality (VR) environments with 65 participants, and demonstrates that olfactory stimuli selectively enhanced EEG responses with positive emotions. Hou et al. [11] found that gamma-band EEG signals can effectively discriminate emotions induced by 13 different odors across binary and five-level valence scales. Meng et al. [12] and Hou et al. [11] established two complementary datasets, EEGDot with 13 odor types and EEGDoc with concentration-varying stimuli, which enable investigation of the effects of odor identity and intensity. Xia et al. [13] synchronously recorded EEG and electronic nose signals for robust cross-subject odor preference assessment, while Guo et al. [14] and Xia et al. [15] developed dual-labeled datasets and annotated food odor responses with both material and affective attributes.

Despite recent progress, olfactory-based emotion recognition still presents two major limitations. First, most existing datasets mainly rely on EEG and neglect ECG and PPG signals, which restricts the modelling of cross-modal correlations. Second, existing methods mostly evaluate emotions from the emotional valence dimension rather than analysing within the arousal valence space, which limits the fine-grained representation of emotions. Therefore, the development of olfactory emotion computing still lags behind the multimodal framework of visual evoked potentials. To bridge this gap, it is necessary to study the multimodal physiological responses triggered by odors and develop models that can effectively integrate heterogeneous signals.

### B. Vision-Induced Multimodal Emotion Recognition Method

Given the scarcity of research on multimodal emotion recognition induced by olfaction, this paper uses the multimodal emotion recognition method induced by vision as an example to introduce and summarise existing related methods.

Recently, graph-based methods have evolved to model complex spatial dependencies in physiological signals. Jia et al. [15] employed attribute graphs for channel correlation modelling, while Zhang et al. [17] introduced group sparse canonical correlation analysis to exploit structural information across EEG and eye movement features.

Recent research has focused on adaptive topological learning to address the issue of inter-subject variability. Huang et al. [18] proposed GJFusion, which that models EEG and peripheral signals as subgraphs and employs Gumbel-Softmax sampling for adaptive channel-level matching. To further address multimodal heterogeneity, Huang et al. [19] introduced CMMGD, which decomposes the graph into consistent and inconsistent subgraphs via correlation-driven separation. In addition to spatial modelling, hybrid architectures have also emerged to capture the temporal dynamics of non-stationary physiological signals. Cheng et al. [20] cascaded a multi-scale dynamic convolutional neural network (CNN) with a gated recurrent unit (GRU) to jointly model local patterns and long-range dependencies, while Wang et al. [22] designed an attention-enhanced 1DCNN-GRU framework for dynamic time-domain alignment between EEG and ECG streams. Li et al. [23] explicitly incorporated 3D topological priors into spatial-temporal transformers for VR-based emotion recognition. Beyond feature extraction, cross-modal alignment strategies have been developed to bridge the semantic gap between physiological signals and emotional labels. Cheng et al. [24] proposed DISD-Net with dynamic interaction modules and self-distillation for EEG-facial fusion, while Zhang et al. [25] proposed visual-to-EEG knowledge distillation to transfer semantic supervision from discriminative modalities. For label-scarce scenarios, Sun et al. [26] constructed hierarchical contrastive masked autoencoders for self-supervised representation learning.

Despite significant progress, current methods still suffer from the following limitations. First, local modelling overlooks global frequency-domain features. Second, the delay inherent in peripheral physiological signals relative to EEG signals is not considered during feature alignment. Finally, conventional splicing and fusion methods ignore the issues of correlation and interference between modalities. To address these limitations, we propose STF-HFNet, which integrates frequency-domain modelling, reciprocal guidance alignment, and hybrid collaborative fusion to ensure robust learning of representations for heterogeneous signals.

## III. Data Collection

This section is organised as follows. First, we introduce the process of odor selection and screening. Next, we describe the multimodal data acquisition platform and experimental setup. Then, we detail the experimental protocol. Finally, we outline the data processing pipeline.

### A. Odor Selection

TABLE I
ODOR AFFECTIVE EVALUATION FORM

| Arousal | Valence |
|---|---|
| 1: Very Weak | 1: Disgust |
| 2: Weak | 2: Dislike |
| 3: Moderate | 3: Neutral |
| 4: Perceptible | 4: Pleasant |
| 5: Clearly Perceptible | 5: Like |

The AMIGOS dataset [27] is primarily based on visual stimuli. In contrast, multimodal datasets triggered by olfactory stimuli remain scarce, and prior studies mostly employ one-dimensional emotion assessment, which makes it difficult to comprehensively characterise participants' emotional experiences. To address this issue, this section introduces the olfactory multimodal dataset we constructed in a real-world setting and provides detailed information on participants' age distribution, experimental environment, and data collection protocol.

The effectiveness of emotion elicitation is directly determined by the selection of stimulus odors. In this paper, 30 odors were preliminarily selected as emotion elicitation materials. Given the limitations of the traditional Self-Assessment Manikin (SAM) [28] in olfactory evaluation, we adapted the scale to better suit the characteristics of olfactory stimuli (the adapted scale is shown in Table I). We selected odors based on the following criteria. First, several odors with proven positive mood-regulating effects were selected from the literature [31] [32] [37], such as bergamot, lemon, rose, jasmine, lavender, and sweet orange. Second, to enrich the diversity of odor samples, we collaborated with a perfumer to create three custom fragrances: Fragrance 1 is a blend of fruity, floral, and woody notes; Fragrance 2 features floral notes and ambergris; and Fragrance 3 is a complex combination of woody and fruity scents. We also selected some negative odors, including civet, castoreum, hyraceum, pepper, galbanum ester, galaxolide, and mint. After consulting with the perfumer, we also chose some neutral odors, such as carrot seed, Tieguanyin tea, forest, myrrh, vetiver, quince, and ocean. To ensure the integrity and balance of the experimental design across emotional dimensions, we similarly incorporated odors with negative emotion-inducing effects (five essential oils with distinct scent profiles, coded as T&T A10-4.0, B10-4.5, C10-5.0, D10-4.5, and E10-5.0.) based on the literature [38] [39] [40].

## B. Odor Screening

TABLE II
GENDER DISTRIBUTION OF VOLUNTEERS IN ODOR TESTS

| Age \ Gender | Female | Male |
|---|---|---|
| 18-25 | 41 | 49 |
| 26-30 | 10 | 8 |
| 31-40 | 1 | 3 |
| 41-50 | 2 | 6 |
| 50+ | 2 | 6 |

To determine the emotional distribution elicited by these odors, we recruited 128 volunteers (who were recruited for the odor screening and did not participate in subsequent data collection) and asked them to rate 30 odors on an arousal-value space. The gender distribution of all participants is shown in Table II. Based on these ratings, each odor was mapped to its corresponding quadrant in the valence-arousal space, as illustrated in Figure. 1. Experimental results indicate that, compared with visual stimuli, olfactory stimuli induce significantly less pronounced emotional changes, particularly in the low arousal-high valence (LAHV) quadrant. In the LAHV quadrant, only a small number of odor samples are distributed in the boundary regions and thus lack typicality. Given this limitation, we ultimately screened only three categories with more distinct emotional characteristics for analysis: high valence-high arousal (HAHV); low valence-high arousal (HALV); and moderate valence-moderate arousal (MAMV). Ultimately, 10 odors were screened, including four MAMV odors (carrot seed essential oil, lemon essential oil, T&T B, and T&T E), three HAHV odors (jasmine essential oil, rose essential oil, and lavender essential oil), and three HALV odors (castoreum essential oil, civet essential oil, and hyraceum essential oil). Screening criteria were based on several classic odors known for their positive effects on emotional recognition, as well as those positioned farther from the origin point in the valence-arousal space.

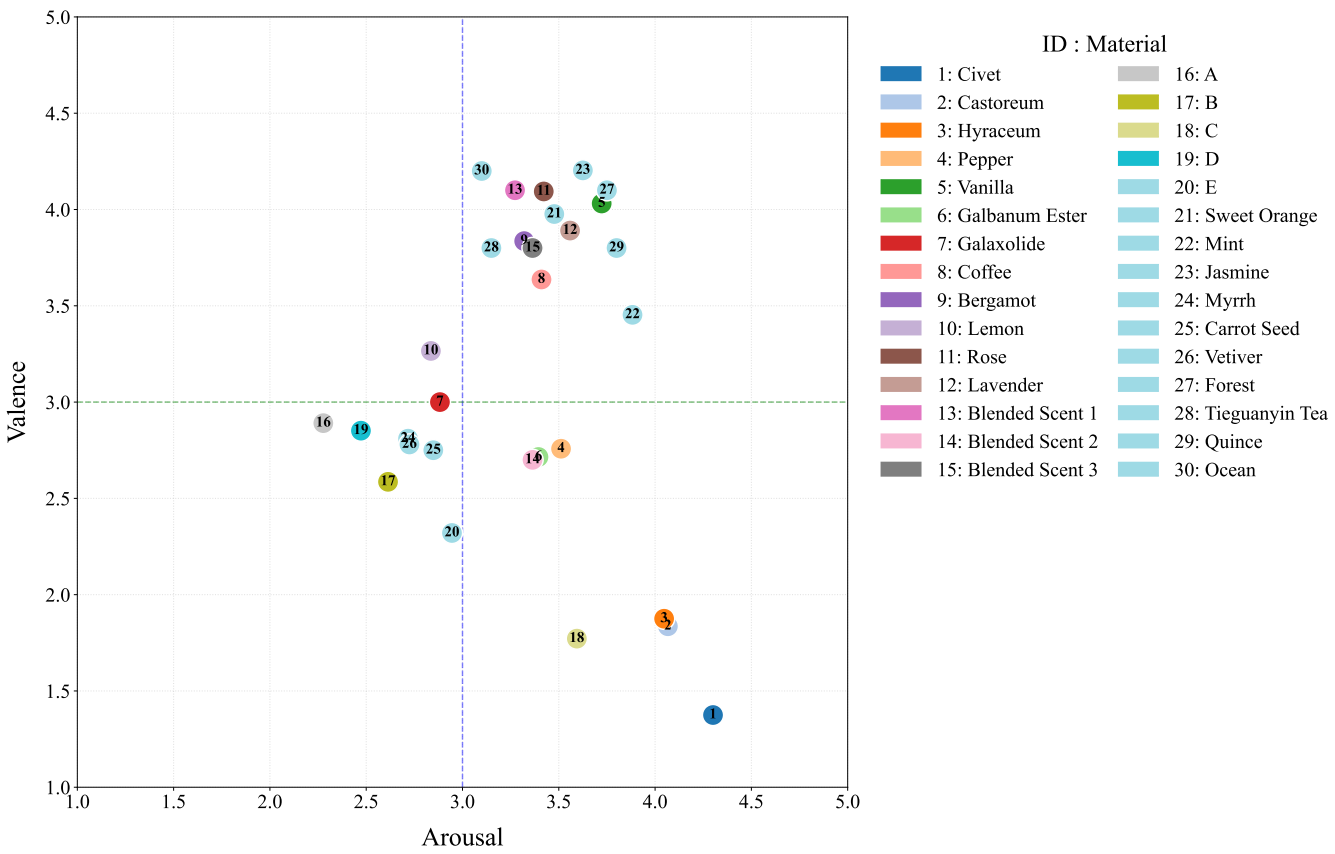

Figure 1: Distribution of Mean Odor Perception Ratings.

## C. Multimodal Data Acquisition Platform and Environment

To ensure that data acquisition aligns with real-world home-use scenarios, portable wearable devices were employed for signal recording in this study. EEG signals were collected using an Emotiv 14-channel wireless EEG headset device, and ECG and PPG signals using a Maxim Integrated MAX86150 evaluation board. As shown in Fig. 2, the MAX86150 module integrates dry electrodes for single-lead ECG measurement and a dual-wavelength pulse oximeter sensor equipped with red and infrared light channels. Both devices communicate wirelessly with the host via Bluetooth. To minimise power line interference, all charging cables were disconnected during the experiment, and the devices were powered entirely by batteries. EEG signals were recorded at 128 Hz and while ECG and pulse oximetry signals at 200 Hz.

**Materials and Procedure:** In this experiment, we selected the 10 odor types listed in Table III to elicit multimodal physiological responses in the participants. To ensure consistent odor intensity, we placed 5 mL of each essential oil into a 125 mL sampling bottle and let it stand for 5 minutes to ensure complete volatilisation.

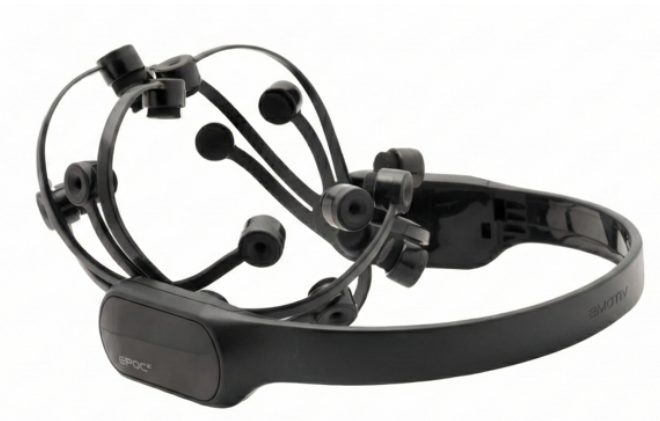

(a) Emotiv Device

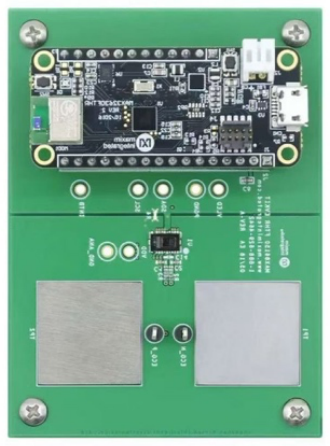

(b) MAX86150 Device

Figure 2: Physiological Signal Acquisition Devices.

**Subject and Environmental Controls:** Prior to the experiment, participants were instructed to fast for 2 hours and abstain from alcohol for 24 hours. They were also required to remove all personal electronic devices. To minimise auditory interference, data collection took place in a soundproof room that was well ventilated, maintained at a constant temperature of 25°C, and free from strong electromagnetic interference.

TABLE III
DETAILED INFORMATION ON THE 10 EXPERIMENTAL ESSENTIAL OILS

| Screened Odors | Main Components | Origin |
|---|---|---|
| Carrot seed essential oil | Carotenol | China |
| Castoreum essential oil | 4-Ethylphenol | Switzerland |
| Civette essential Oil | The macrocyclic musk molecule civetone | Switzerland |
| Hyraceum Essential Oil | Phenylacetic Acid, Skatole | Switzerland |
| Jasmine essential oil | Benzyl Acetate | China |
| Rose essential oil | Citronellol, Geraniol | China |
| Lavender essential oil | Linalyl Acetate, Linalool | China |
| Lemon essential oil | Limonene, Citral | China |
| T&T B | Methylcyclopentenolone | Japan |
| T&T E | Skatole | Japan |

**Experimental Protocol:** After each data acquisition session, subjects were required to leave the room and rest for 5 minutes. This interval reduced fatigue and eliminated olfactory adaptation as well as residual effects from the previous odor.

## D. Design of Olfactory Stimulator

We developed a multi-channel high-precision olfactory stimulation system to ensure spatiotemporal accuracy, as illustrated in Fig. 3. This system precisely controls odor release and flow rate to guarantee consistent stimulus delivery. A semi-open configuration—with the nozzle positioned 3-4 cm from the nasal cavity—effectively suppresses trigeminal tactile artifacts and mechanical airflow noise. By supporting flexible adjustments of parameters such as duration, interval, and flow rate, the system provides robust hardware for acquiring high-quality olfactory-induced multimodal physiological data.

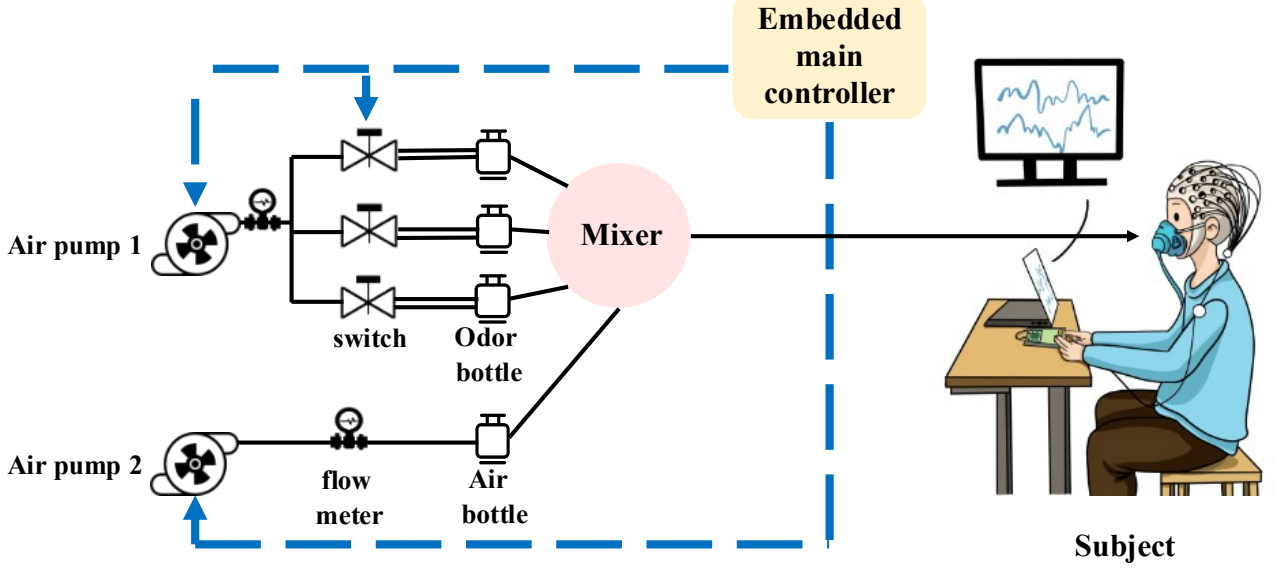

Figure 3: Principle Diagram of Olfactory Stimulator.

## E. Ethics Statement

This study was conducted in accordance with the ethical principles outlined in the Declaration of Helsinki and was approved by the Ethics Committee of Tianjin University (TJUE2026-H-S-059). All subjects were fully informed of the research objectives, data collection procedures, and potential risks prior to the experiment. They voluntarily provided written informed consent and were informed that they had the right to withdraw from the study at any time without penalty or adverse consequences.

## F. Data Collection Protocol

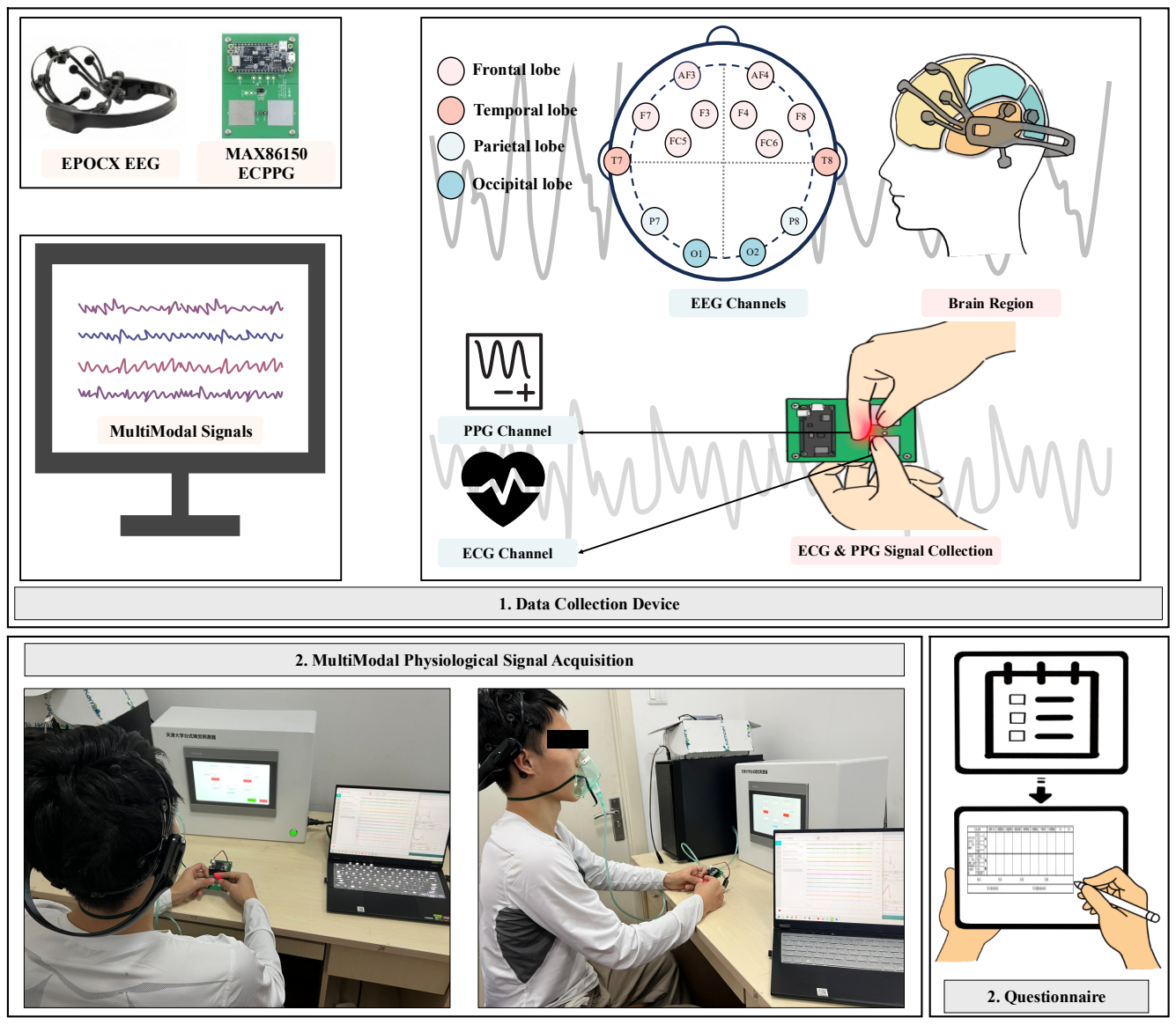

Figure 4: Data Collection Process for Each Subject.

Fig. 4 illustrates the actual collection process of the All-age Multimodal Olfactory Response & Emotion (AMORE) dataset, which includes both a resting-state phase and an olfactory stimulation acquisition phase. A total of 113 subjects were recruited for this study to collect EEG, ECG, and PPG signals. Where 2 subjects were excluded due to poor signal quality, and data from 111 subjects were ultimately included in the analysis. Among them, 60 were male and 51 were female. The age distribution was as follows, 59 subjects aged 18–25, 38 subjects aged 26–30, 6 subjects aged 31–50, and 8 subjects aged 51–65. As shown in Fig. 5, the stimulation-phase consists of 10 trial blocks, each comprising 15 seconds of odor stimulation followed by 10 seconds of rest and repeated for 10 cycles. The rest period helps minimise potential odor habituation, which ensures that each trial captures an independent physiological response to the odor stimulus.

**Resting State Test.** Subjects first sit quietly for 3 min to stabilise their respiratory and physiological states. They then put on the Emotiv EEG headset and the olfactory mask. Next, subjects placed both index fingers on the dry ECG electrodes and covered the PPG sensor (equipped with LED and photodetector) with their right middle finger.

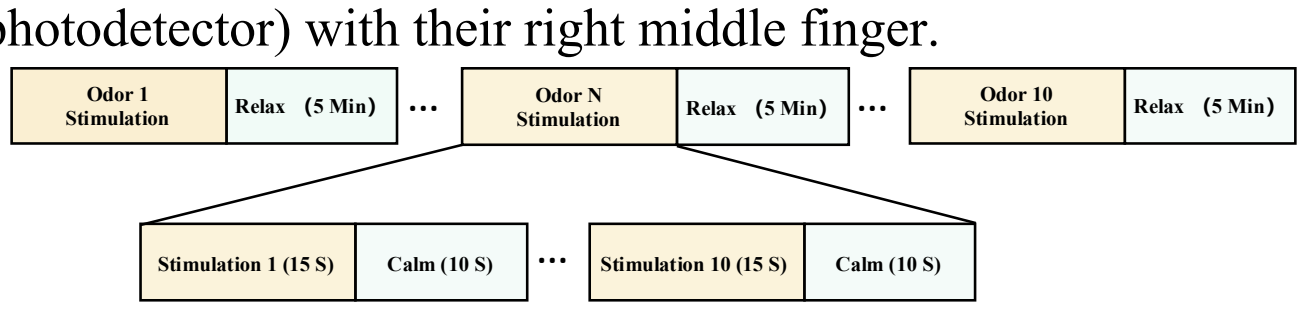

Figure 5: Olfactory Stimulation Process Diagram.

**Olfactory Stimulation Test:** Subjects maintained the same posture and procedure as in the Resting state test to avoid motion artifacts. Fig. 5 illustrates the experimental setup and posture requirements. Once baseline stability of the multimodal signals (EEG, ECG, PPG) was confirmed, the experiment's coordinator initiated olfactory stimulation and recorded the physiological data. The testing procedure (as shown in Fig. 5) is as follows. One odor was selected for 15 seconds of stimulation, followed by a 10-second rest period, with this cycle repeated 10 times. A total of 10 different odors were used in the test.

## G. Data Preprocessing

For each subject, multimodal signals were recorded under 10 odor conditions, with 10 trials per condition. Only the effective stimulation intervals were retained, and the signals were segmented into non-overlapping 5-second windows, which yielded three samples per trial.

EEG preprocessing: Raw EEG signals were processed using the EEGLab toolbox. Fourteen channels corresponding to the 10–20 system locations were retained. The signals were band-pass filtered (0.1–40 Hz) and notch-filtered (48–52 Hz) [29], re-referenced to the common average, and cleaned using independent component analysis-based artifact removal.

ECG preprocessing: ECG polarity was corrected first, followed by a 50 Hz notch filter and a 0.5–60 Hz Butterworth bandpass filter. Residual noise was reduced via wavelet denoising, and baseline wander was suppressed using two-stage median filtering (0.2 s and 0.6 s windows).

PPG preprocessing: PPG signals were bandpass filtered (0.5–10 Hz, 4th-order Butterworth) and smoothed with a 1-second moving average. We further applied NeuroKit2 [30] for artifact cleaning and quality control, and employed two-stage median filtering to eliminate baseline drift.

We also used data from the AMIGOS dataset [27], which captures visually evoked emotional changes. This subset comprises 40 healthy participants (13 females) aged 21-40 years. During the experiment, all subjects watched 16 short video clips individually. These clips, with duration ranging from 51 to 150 seconds, were screened to elicit emotional responses covering all four quadrants of the valence-arousal space. The experiment collected multimodal neurophysiological signals using wearable devices, including a 14-channel EEG, ECG, and galvanic skin response (GSR).

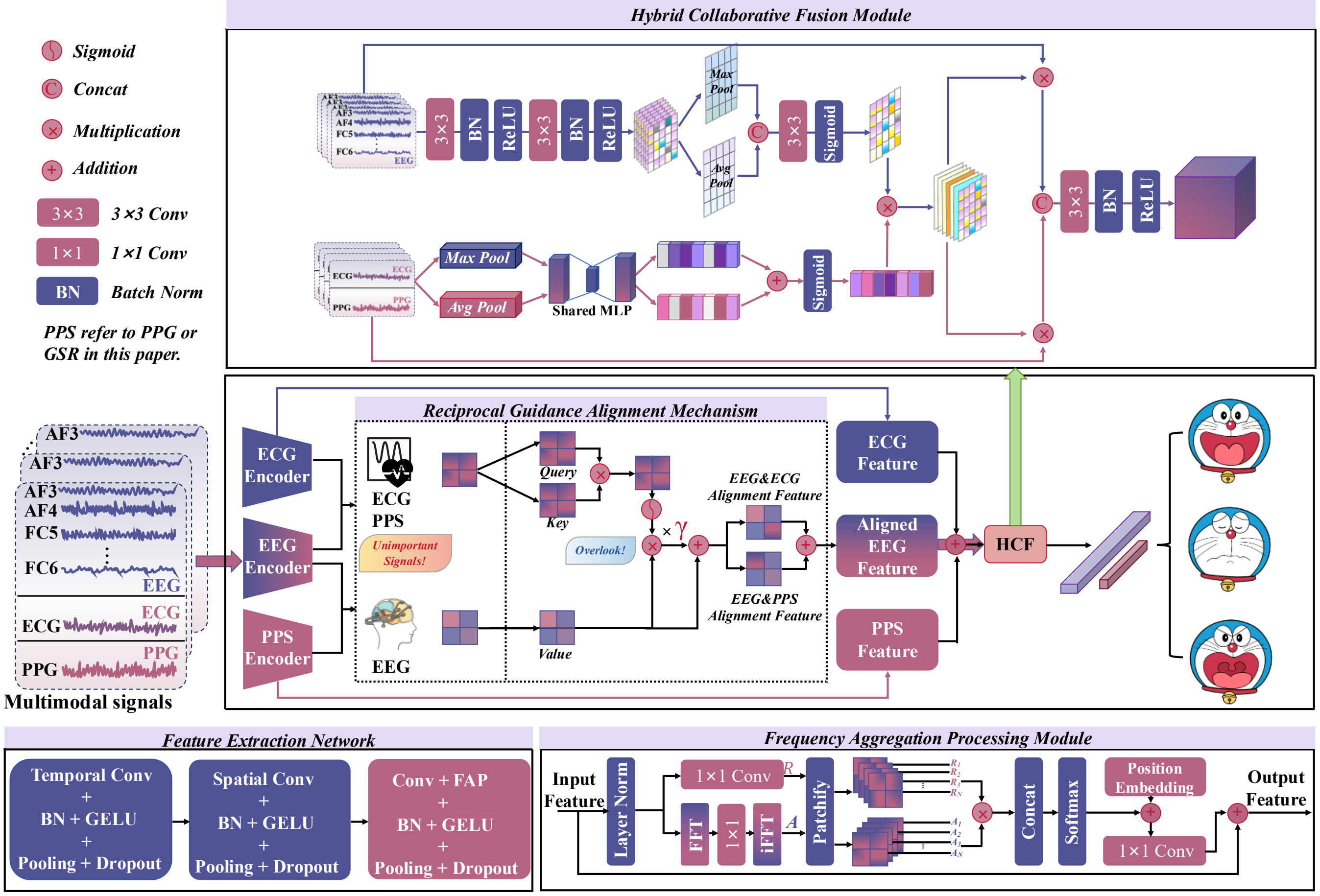

Figure 6: STF-HFNet Model Framework (Peripheral physiological signals (PPS) refer to GSR in the AMIGOS dataset and PPG in the AMORE dataset). Among them, the feature extraction network refers to the encoder of three modalities, FAP refers to the frequency aggregation processing module, and HCF refers to the hybrid collaborative fusion module.

Our evaluation incorporates three modalities—EEG, ECG, and GSR—to perform a 4-class emotion recognition task based on valence and arousal dimensions. Subjects rated their emotional states on a scale of 0 to 9. According to the literature [36], we adopted a thresholding strategy for label generation: Ratings above 5.5 were categorised as high, those below 4.5 as low, and ratings that fell between these thresholds were considered neutral and excluded from the dataset. Consequently, the emotional states were classified into four distinct quadrants: high valence-low arousal (HVLA), high valence-high arousal (HVHA), low valence-high arousal (LVHA), and low valence-low arousal (LVLA).

## IV. Methodology

In this section, we elaborate on the proposed spatiotemporal-frequency hybrid fusion network (STF-HFNet), as illustrated in Fig. 6. The architecture of STF-HFNet is directly driven by the need to overcome the key challenges of signal non-stationarity, physiological asynchrony, and semantic heterogeneity. Multimodal physiological signals exhibit time-varying frequency dynamics that conventional spatiotemporal methods fail to capture. The FAP module therefore performs FFT-based frequency reweighting combined with multi-head patch decomposition to obtain stable frequency-aware representations for each modality. Built on these stabilised representations, the central and peripheral streams still suffer from inherent physiological propagation delays; the RGA mechanism thus leverages the structure information of peripheral signals as queries and keys to recalibrate EEG features and generates dynamically temporally coherent cross-modal representations. Finally, although the aligned features are now temporally synchronised, EEG and ECG/PPG remain semantically heterogeneous in terms of channel dimensionality and information granularity. The HCF module therefore constructs a hybrid attention guidance map that integrates EEG channel semantics with peripheral spatial saliency and selectively enhances complementary cues to generate a unified discriminative representation.

Our problem definition is as follows. Given a multimodal physiological dataset $D = \left\{\left(S^{(i)}, y^{(i)}\right)_{N_i=1}(N_i = 1,2,\dots,C)\right\}$ collected during olfactory stimulation or from a public dataset, where $1,2,\dots,C$ represent the emotion label, $S^{(i)} = \{S_{eeg}{}^{(i)},$ $S_{ecg}{}^{(i)}, S_{pps}{}^{(i)}\}$ represents a signal window (including EEG, ECG, and PPS (PPG or GSR)). $S^{(i)} \in \mathbb{R}^{M\times N}$, where $M$ denotes the number of sensor channels, and $N$ denotes the length of the time windows. And $y^{(i)} \in \{1,2,\dots,C\}$ denotes the emotion label, our objective is to learn a mapping model that can accurately predict emotional states.

### A. Frequency Aggregation Processing Module

Physiological signals are intrinsically characterised by frequency-domain properties: EEG frequency bands [33] (delta, theta, alpha, beta, gamma) encode cognitive states, while ECG and PPG frequency components reflect cardiac rhythms. Traditional CNNs operate exclusively in the spatiotemporal domain, and thus fail to explicitly capture these diagnostic frequency bands [29]. We propose the FAP module to mitigate the limitation whereby ignoring the frequency-domain structure makes it difficult to model long-range dependencies in non-stationary physiological signals: FAP performs FFT-based spectral reweighting and an iFFT projection to capture global frequency-domain information across the entire time series.

After feature extraction, the original signal is converted into a signal feature map $X \in \mathbb{R}^{N\times C\times H\times W}$, where $N$ denotes the batch size, $C$ represents the number of feature channels, and $H$ and $W$ are the height and width of the feature map, respectively. Given the input features $X$, we perform FFT to transform the features into the frequency domain, where a $1\times 1$ convolution is employed to learn global features and implement spectral reweighting. Subsequently, we use iFFT to project the modulated spectrum back to the temporal domain. According to the spectral convolution theorem [34] [35], our method is capable of capturing global receptive fields across the entire time-series signals. Following the above analysis, we convert the time-domain features into frequency-domain features using Eq. (1).

$$\hat{A} = \mathcal{P}\left(\mathcal{F}\left(LN(X^{(i)})\right)\right), \tag{1}$$

where $\mathcal{P}(\cdot)$ denotes the $1\times 1$ convolution function used to re-weight the frequency-domain physiological feature, $\mathcal{F}(\cdot)$ represents the FFT, and $LN(\cdot)$ is the layer normalisation function. Then, we employ the iFFT function to convert the $\hat{A}$ back to the temporal domain, as shown in Eq. (2).

$$A = i\mathcal{F}(\hat{A}), \tag{2}$$

$i\mathcal{F}(\hat{A})$ denotes the iFFT operation. According to Eq. (2), we can obtain the frequency aggregation feature $A$. Meanwhile, a parallel $1\times 1$ convolution operation generates original temporal domain represent embeddings $R$ that preserve spatial-temporal information, which enables the FAP module to integrate both frequency-domain and time-domain representations.

Then, to capture multi-scale physiological patterns, we reshape both $A$ and $R$ from $\mathbb{R}^{H\times W\times C}$ to a multi-head format $\text{Head} \times \frac{C}{\text{Head}} \times M$, where $M = H\times W$. We then partition features into $L$ non-overlapping temporal-spatial patches, yielding patch sets $[A_1, A_2, \dots, A_N]$ and $[R_1, R_2, \dots, R_N]$. Finally, we perform element-wise multiplication within each patch to generate the temporal and frequency aggregation features. This patch decomposition enables the network to simultaneously model temporal-frequency correlations (e.g., sustained EEG signal oscillations and heart rate variability) and localised transient events (e.g., changes in heart rate or EEG rhythm) from different perspectives. However, the FFT does not consider local spatial information in the time domain, which results in the loss of critical local waveform morphology. To address this issue, we introduce positional encoding, which explicitly injects local temporal features through a lightweight convolution, as shown in Eq. (3).

$$FAP = Q\left(\text{Softmax}\left(\underset{1\le n\le L}{\text{Concat}}(A_n \otimes R_n)\right) + E(X^{(i)})\right) + X^{(i)} \tag{3}$$

Here, $E(\cdot)$ represents the positional embedding, which is implemented via a $3\times 3$ convolution operation, $\otimes$ denotes element-wise multiplication, and $Q(\cdot)$ is the projection function implemented by a $1\times 1$ convolution. To prevent gradient vanishing, we add the frequency-domain enhanced features to the original features to obtain the final output.

### B. Reciprocal Guidance Alignment Mechanism

The physiological manifestation of emotion is characterised by cognitive arousal in the EEG, coupled with autonomic responses exhibited in PPS signals, such as changes in heart rate and vascular tone. Conventional concatenation fusion methods, which process each modality independently, fail to learn a unified representation of these synergistic physiological patterns. To address this limitation, we propose the RGA mechanism to explicitly model these inter-modal dependencies. The RGA mechanism enables the internal feature structure of one modality to dynamically guide the recalibration of another modality's features, and thereby learn a holistic representation that captures emotion-specific coordination between the central and autonomic nervous systems.

The RGA mechanism takes dominant features $X_{eeg}$ and auxiliary features $X_{aux}$ ($X_{ecg}$ or $X_{pps}$) as inputs. Specifically, in the RGA mechanism, both $Q_{aux}$ and $K_{aux}$ are derived from the $X_{aux}$ by performing transposition and multiplication of these components, the model learns information from multiple feature map perspectives and dynamically adjusts the feature of the EEG modality. Unlike traditional self-attention mechanisms, the RGA mechanism only uses $Q_{aux}$ and $K_{aux}$ from the same signal source and does not introduce a Value component. This design is intended to maintain the signal consistency and accuracy of guidance during feature calibration. Using the same signal for $Q_{aux}$ and $K_{aux}$ ensures more precise guidance for the EEG modality and avoids interference and inconsistency that could arise from cross-modal signals. This improves the alignment and fusion accuracy of central and peripheral physiological responses. We capture this synergistic relationship through Eq. (4).

$$G_{ij} = \frac{exp\left(Q_{aux}^{i}, K_{aux}^{j}\right)}{\sum_{i=1}^{C} exp\left(Q_{aux}^{i}, K_{aux}^{j}\right)}. \tag{4}$$

Here, $G_{ij}$ quantifies the strength of the contribution of the $i^{\text{th}}$ channel to modulating the $j^{\text{th}}$ channel, and captures internal structural dependencies within the EEG feature space. $C$ denotes the number of channels, and $exp\,(\cdot,\cdot)$ serves as a similarity function between two channel representations. $Q_{aux} \in \mathbb{R}^{N\times C\times T}$ and $K_{aux} \in \mathbb{R}^{N\times T\times C}$ are reshaped from $X_{aux}$ by flattening spatiotemporal dimensions, where $T = H \times W$.

EEG feature channels encode multi-frequency and spatial patterns, with their emotion-discriminative importance varying with autonomic states. We dynamically reorganise these channels using the guidance matrix $G$ from auxiliary modalities (e.g., ECG/PPG/GSR) , as shown in Eq. (5).

$$Z_j = \gamma \sum_{i=1}^{C} \left(G_{ij} \cdot V_{dom}\right) + X_{eeg}^{j}, \tag{5}$$

Where $V_{dom} \in \mathbb{R}^{B\times C\times E}$ is reshaped from $X_{eeg}$. This adaptive channel mixing operation reweights EEG features based on autonomic co-activation patterns. $\gamma$ is initialised to zero and adaptively learns the cross-modal guidance strength. Specifically, $\gamma$ automatically acquires instance-wise fusion strength, acting as an adaptive gate that modulates the guidance effect based on input patterns. We extract $Q_{aux}$ (Query) and $K_{aux}$ (Key) from $X_{aux}$ rather than $X_{eeg}$, because although peripheral signals are low-dimensional, they exhibit stable and interpretable inter-feature dependencies that directly reflect the body's autonomic responses. By computing self-attention on $X_{aux}$, we obtain a robust structural prior $G_{ij}$, which is used to guide the recalibration of high-dimensional EEG channels ($V_{dom}$) and effectively filter out neural activity unrelated to emotional responses.

We adopt EEG as the primary modality and employ a many-to-one guidance strategy to obtain complementary information from autonomic signals (ECG, PPS), as shown in Eq. (6).

$$RGA = \frac{1}{2}\left(Z\left(X_{eeg}, X_{ecg}\right) + Z\left(X_{eeg}, X_{pps}\right)\right). \tag{6}$$

This allows the neural features of EEG to simultaneously receive cross-modal guidance from both ECG and PPS, which complementarily recalibrate the cognitive-affective representations of EEG. As auxiliary modalities, ECG and PPS provide autonomic-state priors through their co-activation patterns, guiding the EEG modality to dynamically select emotion-relevant frequency bands and spatial patterns. In turn, this enhances the discriminative capability of the primary EEG modality.

### C. Hybrid Collaborative Fusion

Following feature extraction and multimodal alignment, the critical challenge lies in effectively fusing heterogeneous physiological representations into a unified and discriminative feature. Since the features are derived from different physiological signals, the primary stream (EEG signal $X_{eeg}$) contains rich high-level semantic features but may lack distinct temporally localised information. In contrast, the auxiliary streams (ECG and PPS signals $X_{aux}$) possess obvious localised variation characteristics, which yield inconsistent semantic and structural features between the primary and auxiliary stream. To address this issue, we construct the HCF module, which achieves feature fusion by designing a hybrid attention guidance map.

First, to enhance the consistency of the final fused representation, a heterogeneous map $HM \in\cdot \mathbb{R}^{C\times H\times W}$ is designed based on the semantic features of $X_{eeg}$ and the structural features of $X_{aux}$.

To preserve the complementary information across different modalities during the fusion process, we use the channel attention map generated by $X_{eeg}$ to capture semantic consistency and the spatial attention map generated by $X_{aux}$ to capture structural consistency. We then multiply these two maps element-wise to obtain $HM$, which guides the joint fusion of semantic and structural features. The generation process for $HM$ is defined as follows:

$$F_{aux} = ReLU\left(BN\left(Conv_{3\times3}\left(ReLU\left(BN\left(Conv_{3\times3}(X_{aux})\right)\right)\right)\right)\right), \tag{7}$$

$$W_{aux} = Sigmoid\left(Conv_{3\times3}\left[P_{max}(X_{aux}); P_{avg}(X_{aux})\right]\right), \tag{8}$$

$$W_{dom} = Sigmoid\left(FC\left(P_{avg}(X_{eeg})\right) + FC\left(P_{max}(X_{eeg})\right)\right), \tag{9}$$

$$HM = W_{aux} \odot W_{dom}. \tag{10}$$

Where $W_{aux}$ and $W_{dom}$ are the spatial attention weight map and the channel attention weight map, respectively. $P_{max}$ and $P_{avg}$ denote max pooling and average pooling operations across the corresponding dimensions, and $[\cdot\,;\cdot]$ denotes the concatenation operation.

Subsequently, the generated guidance map $HM$ facilitates a bidirectional, synergistic refinement between the two streams. On one hand, $X_{eeg}$ (EEG features) is modulated by $HM$ to prioritise its high-level features within the temporal regions identified as salient by the spatial attention component (derived from $X_{aux}$). This step effectively leverages the structural guidance from the auxiliary stream to anchor the semantic features of the primary stream (EEG) in critical time windows, which ensures the alignment of central nervous system responses with autonomic regulation. On the other hand, $X_{eeg}$ is concurrently modulated by $HM$ to recalibrate its feature channels, based on the high-level contextual information provided by the channel attention component (derived from $X_{eeg}$ itself). Finally, the reconstructed features are concatenated and passed through a convolution block composed of a convolution layer, a batch normalisation layer, and a ReLU layer and produce the final fused feature maps $F_{out}$. The overall procedure for feature reconstruction and multimodal fusion is formulated in Eq. (11).

$$F_{out} = \mathrm{ReLU}\left(BN\left(\mathrm{Conv}_{3\times3}\left(\left[HM \otimes X_{eeg};\ HM \otimes X_{aux}\right]\right)\right)\right) \tag{11}$$

where $\otimes$ denotes element-wise multiplication. The output $F_{out}$ is then sent to a global average pooling layer to obtain a fixed-length vector for the final classification stage.

## V. Experiments

### A. Platform and Hyperparameters

All models were trained and evaluated on a hardware configuration consisting of four NVIDIA RTX 4090D 24 GB GPUs, two AMD EPYC 7642 CPUs, and 256 GB RAM. The experiments were conducted using the PyTorch deep learning framework, and hyperparameters are presented in Table IV.

TABLE IV
HYPERPARAMETERS

| Parameters | AMIGOS | AMORE |
|---|---|---|
| Temporal Conv Kernel | (1,25) | (1,25) |
| Spatial Conv Kernel | (14,1)/(2,1)/(1,1) | (14,1)/(1,1)/(1,1) |
| Fusion Conv Kernel | (3,3) | (3,3) |
| Dropout | 0.5 | 0.5 |
| Optimiser | Adam | Adam |
| L2 Regularisation | 1e-2 | 1e-2 |
| Epoch | 200 | 200 |
| Learning Rate | 3e-4 | 3e-4 |
| Train : Test | 8 : 2 | 8 : 2 |
| Batch Size | 256 | 256 |

### B. Comparative Experiments

TABLE V
COMPARATIVE EXPERIMENTAL RESULTS FOR TWO DATASETS

| Data set | Model | Accuracy (%) | F1-score (%) | Kappa (%) | PAM (%) |
|---|---|---|---|---|---|
| AMIGOS | TSMMF [41] | 72.49 | 71.59 | 62.05 | 54.99 |
| | SleepTransformer [42] | 76.33 | 75.90 | 67.72 | 62.09 |
| | NeuroNet [43] | 67.70 | 67.06 | 55.81 | 50.22 |
| | Husformer [44] | 74.56 | 74.23 | 65.38 | 59.86 |
| | CH-Net [45] | 76.09 | 75.90 | 67.29 | 61.84 |
| | HyperFuseNet [46] | 71.62 | 74.49 | 61.27 | 56.11 |
| | H2 [47] | 77.40 | 77.42 | 69.34 | 64.58 |
| | Rodriguez *et al.**[48] | 79.20 | 79.60 | - | - |
| | OPWA* [36] | 80.07 | - | - | - |
| | **Ours** | 88.34 | 88.27 | 84.02 | 80.14 |
| AMORE | TSMMF [41] | 72.40 | 71.29 | 61.91 | 54.82 |
| | SleepTransformer [42] | 71.91 | 71.57 | 57.29 | 54.38 |
| | NeuroNet [43] | 78.84 | 78.70 | 67.90 | 63.99 |
| | Husformer [44] | 87.33 | 87.21 | 80.76 | 77.11 |
| | CH-Net [45] | 71.45 | 71.17 | 56.64 | 53.69 |
| | HyperFuseNet [46] | 70.45 | 70.16 | 55.08 | 52.54 |
| | H2 [47] | 73.08 | 72.93 | 59.41 | 56.48 |
| | **Ours** | 92.40 | 92.31 | 88.42 | 86.15 |

Where * represent the results reported directly from the references.The best results are colored as follows: first, second, third.

We compared our results with several SOTA methods. Given that there are only four classification tasks in the AMIGOS dataset, we replicated the following open-source multimodal physiological signal classification algorithms: TSMMF [41], SleepTransformer [42], NeuroNet [43], Husformer [44], CH-Net [45], HyperFuseNet [46], and H2 [47].

Table V presents our comparative experimental results. On the AMIGOS dataset, our method surpasses the SOTA methods by 8.27% in Accuracy, 8.67% in F1-score, 14.68% in Kappa, and 15.56% in PAM. Similarly, on our private dataset, our method outperforms the SOTA methods by 5.07% in Accuracy, 5.1% in F1-score, 7.66% in Kappa, and 9.04% in PAM. These results demonstrate the effectiveness of our proposed method in emotion recognition.

Compared with TSMMF [41] and Husformer [44], which mainly rely on symmetric cross-modal transformers for alignment and interaction, our method adopts a staged pipeline that better fits physiological affect signals. TSMMF/Husformer implicitly assumes that the modality features already lie in an alignable space and use bidirectional attention to learn complementarity. However, in emotion recognition, EEG exhibits significant variability in the frequency domain, while ECG/PPG typically reflects local peripheral state changes. Therefore, direct bidirectional interaction will propagate noise and amplify misalignment. To solve this problem, our method follows a stable recalibration gate design. First, the FAP module explicitly models frequency-domain dependencies to stabilise unimodal representations. Second, the RGA module performs peripheral-guided one-way recalibration. It learns guidance information from the co-activation patterns of ECG/PPG signals and uses this guidance to modulate the importance of EEG channels and the frequency band. And then, RGA employ a learnable $\gamma$ adaptively controls the guidance strength per sample to suppress irrelevant noise. Third, the HCF module leverages peripheral saliency to provide temporal localisation, and combines it with EEG semantic attention for gating, which reduces non-informative time periods and strengthens the coordinated representations of brain and peripheral signals. Compared with SleepTransformer [42] (which is sensitive to temporal mismatch) and hypercomplex/global fusion methods (including CH-Net [45], HyperFuseNet [46], and H2 [47]) that are susceptible to interference from weak long-duration signals and noise, our proposed pipeline yields more consistent cross-modal representations and improves overall performance.

### C. Ablation Study

TABLE VI
ABLATION STUDIES (%). BLACK: AMIGOS, GREEN: OURS DATASET

| Method | Accuracy | F1-score | Kappa | PAM |
|---|---|---|---|---|
| STF-HFNet | **92.40&88.34** | **92.31&88.27** | **88.42&84.02** | **86.15&80.14** |
| w/o FAP | 89.40&86.36 | 89.28&86.11 | 83.87&81.30 | 81.20&76.79 |
| w/o RGA | 90.23&86.31 | 90.18&85.95 | 85.11&81.18 | 82.55&76.56 |
| w/o HCF | 87.86&86.01 | 87.77&85.86 | 81.59&80.88 | 78.74&76.60 |
| Baseline + FAP | 86.13&83.31 | 85.99&83.27 | 78.89&77.25 | 75.75&72.78 |
| Baseline + RGA | 86.44&84.87 | 86.22&84.67 | 79.38&79.35 | 76.28&74.98 |
| Baseline + HCF | 88.63&85.44 | 88.51&85.26 | 82.68&80.18 | 79.87&75.91 |
| Baseline | 84.63&82.26 | 84.41&81.81 | 76.60&75.76 | 73.30&70.72 |

To validate the contribution of each component in STF-HFNet, we conducted ablation studies on both datasets by removing FAP, RGA, and HCF individually. As shown in Table VI, removing any single module consistently degrades model performance, which confirms that all three components are indispensable for the model's effectiveness. Specifically, excluding FAP lowers the accuracy, which indicates the importance of frequency-domain global modelling in capturing rhythm-related physiological feature. Removing RGA further impairs performance, and suggests that cross-modal alignment helps mitigate discrepancies between the two modalities and improves fusion quality. The most significant performance drop occurs when HCF is removed; this highlights the fact that high-quality multimodal fusion is the key factor driving the model's performance improvement. Overall, these ablation results demonstrate that the superior performance of STF-HFNet stems from the complementary effects of FAP, RGA, and HCF.

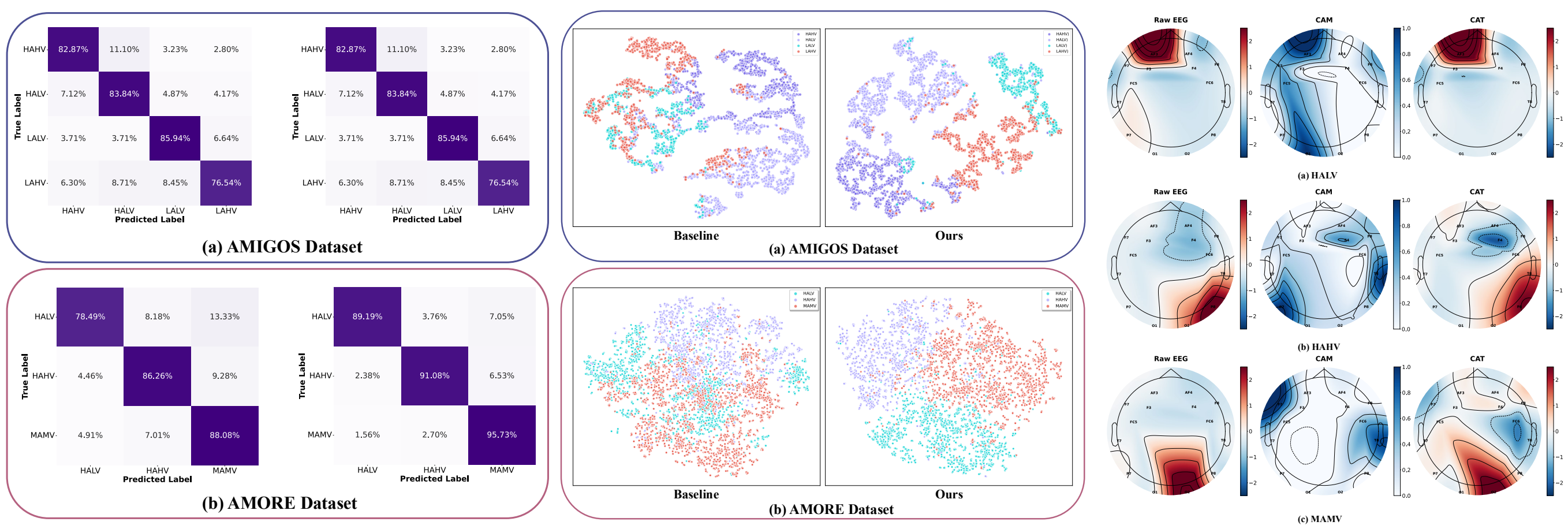

Figure 7: Confusion Matrix on Two Datasets.

Figure 8: t-SNE on Two Datasets.

Figure 9: Brain Topography Visualisation and Analysis.

## D. *Multimodal Ablation Analysis*

To evaluate the contribution of each modality to model performance, we conducted experiments on STF-HFNet under different modality input configurations, with the results shown in Table VII. In the single-modality setting, EEG performed best, primarily because it can directly reflect changes in brain activity related to emotional processing. In contrast, ECG and PPS primarily characterise cardiac electrical activity and peripheral blood volume changes, respectively; therefore, when used alone, their ability to distinguish emotions is relatively limited. Combining EEG with either ECG or PPS significantly improved model performance, which indicates that joint modelling of central and peripheral physiological signals is more effective than relying solely on a single physiological signal. In contrast, the performance gains from combining ECG and PPS were relatively limited, suggesting that improvements in model performance depend more on the complementarity between modalities than on their simple superposition. When all three modalities were used simultaneously, the model achieved the best results. This further illustrates that ECG and PPS provide peripheral physiological information complementary to EEG, thereby enabling more comprehensive characterisation of emotional states.

TABLE VII
MODAL ABLATION ANALYSIS (%). BLACK: AMIGOS, GREEN: OUR DATASET (WHERE PPS REFER TO GSR IN AMIGOS, AND PPG IN OUR DATASET).

| Modality | Accuracy (%) | F1-score (%) | Kappa (%) | PAM (%) |
|---|---|---|---|---|
| EEG | 78.15&79.53 | 77.82&79.41 | 66.61&71.99 | 63.18&66.98 |
| ECG | 58.06&63.23 | 57.69&63.23 | 36.48&50.17 | 35.94&45.75 |
| PPS | 54.16&38.42 | 43.70&26.75 | 25.57&9.90 | 29.70&15.82 |
| EEG + ECG | 88.33&86.13 | 88.23&85.92 | 82.24&80.98 | 79.33&76.66 |
| EEG + PPS | 87.35&84.36 | 87.23&84.24 | 80.71&78.60 | 77.60&74.06 |
| ECG + PPS | 66.00&69.70 | 65.36&69.44 | 48.51&58.62 | 45.06&53.53 |
| STF-HFNet | 92.40&88.34 | 92.31&88.27 | 88.42&84.02 | 86.15&80.14 |

## E. *Visualisation Analysis*

As shown in Fig. 7, the improved model exhibits stronger diagonal values and fewer misclassified samples in the confusion matrix, indicating superior recognition performance. This improvement stems from the synergistic effects of three modules. Specifically, FAP captures discriminative rhythmic patterns through frequency-domain global modelling. RGA aligns cross-modal representations and promotes effective interaction among EEG, ECG, and PPG features. HCF further enhances multimodal fusion through collaborative attention, which enables the model to preserve complementary information from different physiological signals.

As shown in Fig. 8, the t-SNE visualisation further corroborates the above analysis. The feature distribution of the baseline model exhibits obvious class overlap, whereas the improved model forms compact clusters with distinct inter-class boundaries. This indicates that the collaborative effect of the three modules effectively improves intra-class compactness and inter-class separability, and thereby validates the superiority of the proposed method from the perspective of representation learning.

As shown in Fig. 9 (the blue area in the CAM diagram represents the area of network attention; CAM represents class activation, and CAT represents class activation topology; we conducted interpretability analysis based on CAM). To verify the physiological correlation of the features learned by STF-HFNet, we visualised CAM under different emotional states. For HAHV, the model mainly focuses on the left frontal lobe region (AF3, F7, and FC5) and the occipital lobe region (O1). In contrast, for HALV, the attention center significantly shifted to the right frontal lobe region (F4 and AF4) and the temporal parietal lobe region (T8, P7, and P8). Finally, MAMV exhibited a more balanced lateral distribution (F7, T8, and O2). This indicates that STF-HFNet captured features consistent with the neural olfactory paradigm, and thus supports the interpretability of the proposed framework.

## F. *Cross-modal Association Analysis*

Fig. 10(b) visualises the cross-modal correlations between electroencephalogram (EEG) and peripheral signals in response to olfactory stimuli and shows the top 20% of connections (the width of each connection indicates its strength). We observed significant emotion-dependent patterns: The HAHV state primarily activated coupling between prefrontal regions (F3, F4) and peripheral signals, whereas the HALV state enhanced the association between the frontocentral and temporal regions (FC5, FC6, T7) and ECG/PPG signals. This phenomenon reflects the functional differentiation of brain regions based on olfactory value. In contrast, analysis of the AMIGOS dataset (under visual stimulation) in Fig. 10(a) indicates that correlations are primarily concentrated in the occipital (O1, O2) and frontal regions with ECG signals. Unlike visual stimulation, the significant correlated regions under olfactory stimulation are mainly distributed in the frontal and temporal lobes, while occipital correlations are relatively weak.

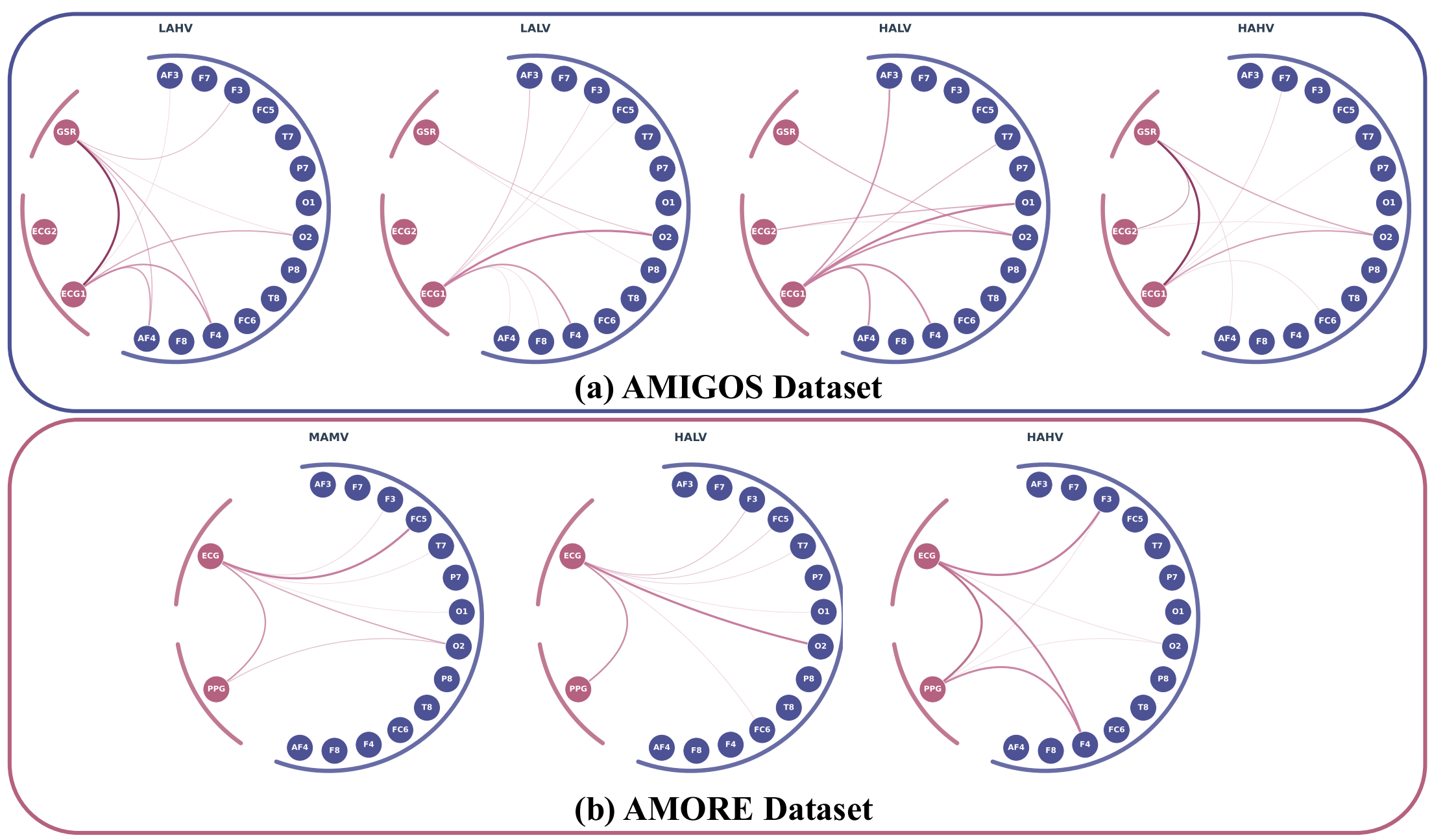

Figure 10: Channel Correlations Analysis.

### G. *Alignment coefficient analysis*

We analysed the alignment effect and found, by visualising the gamma coefficient in the RGA module, that in the AMIGOS dataset (as shown in Fig. 11), the $\gamma$ coefficient tends to stabilise after the 90th epoch. In our private dataset, it stabilises after the 60th epoch. This observation may be attributed to the fact that our private dataset only includes three classification tasks, which renders the learning of feature alignment simpler than the four-classification task of the AMIGOS dataset. We also found that the alignment weights between EEG and ECG were larger, indicating that the signal changes induced by emotional states are mainly concentrated in the EEG and ECG modalities.

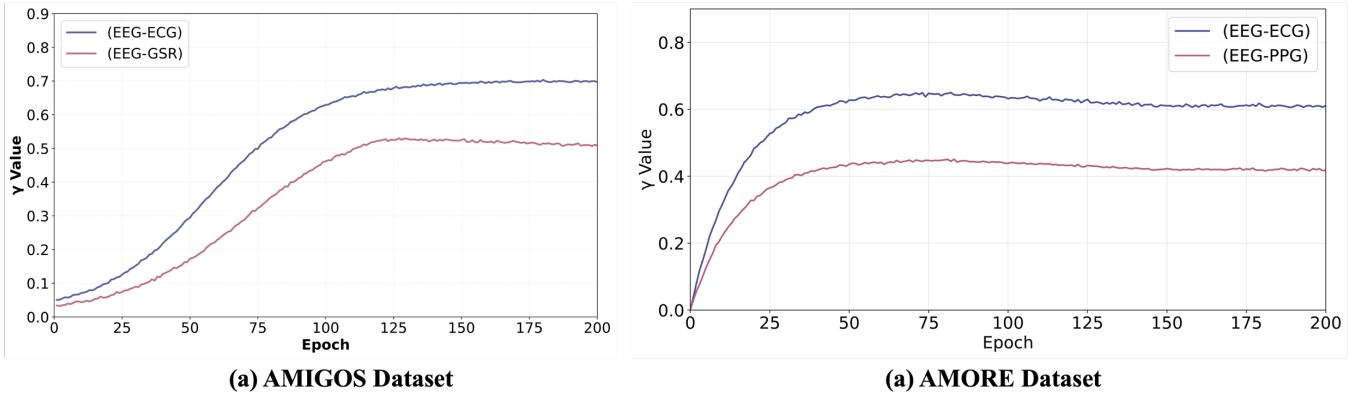

Figure 11: Alignment Coefficient Analysis Experiment.

## VI. Discussion

In this section, we use the AMIGOS dataset as an example to analyse the functional components of each module.

### A. *Internal structure analysis of FAP*

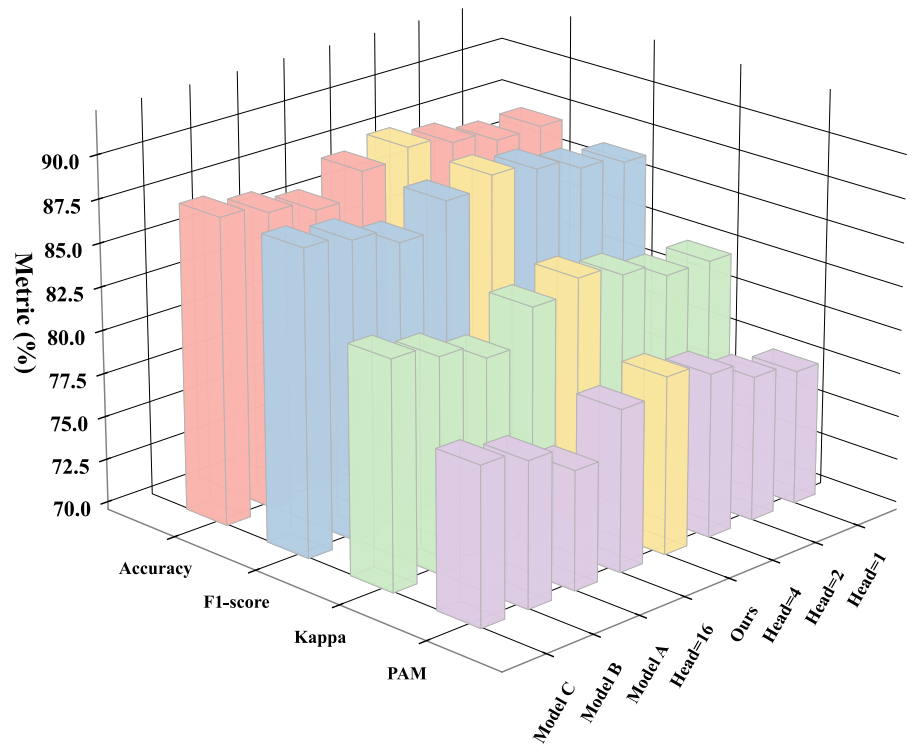

Figure 12: FAP components analysis. Model A (baseline without FAP), Model B (adds spatial-Fourier interaction), Model C (adds patch-based modulation), and our full model (adds multi-head design with position encoding).

As shown in Fig. 12, Model B outperforms Model A, confirming that frequency-domain reweighting effectively captures global spectral dependencies. Model C further enhances multi-scale modelling capabilities. Ultimately, the full model achieves optimal performance, which demonstrates that multi-head decomposition and CPE work synergistically to effectively preserve both global spectral dynamics and local waveform morphology.

Fig. 13 depicts performance scales with the number of heads from 1 to 8. This confirms that parallel subspaces facilitate finer-grained frequency decomposition, which is conducive to capturing distinct physiological bands. When the number of heads is set to 8, the model achieves the highest accuracy; in contrast, setting the number of heads to 16 leads to a slight performance degradation, which is attributed to the over-fragmentation of channel representations.

### B. Ablation Study on Dual-branch RGA Alignment

We conducted ablation experiments on the internal structure of the RGA module and report the results in Table VIII. RGA improves model performance by reducing cross-modal representational mismatch. Without RGA, multimodal fusion is mainly static and occurs at the final stage, in which modality-specific correlations and noise are not explicitly aligned; as a result, auxiliary features interfere with the backbone network and cause representation drift, thereby weakening feature separability. RGA performs conditional dynamic recalibration: It learns channel dependencies in the auxiliary branch and uses this information to reweight the backbone responses, serving as an auxiliary-driven adaptive re-parameterisation mechanism. The gated residual starts near zero, enabling a smooth warm-up process from “no alignment” to “effective alignment”, which stabilises the optimisation process and mitigates early-stage negative transfer. Ablations results show that either single-branch alignment consistently outperforms the model without RGA, and dual-branch alignment achieves optimal performance. This indicates that the two branches provide complementary alignment signals.

TABLE VIII
ABLATION STUDY ON DUAL-BRANCH RGA ALIGNMENT.

| Model | Accuracy (%) | F1-score (%) | Kappa (%) | PAM (%) |
|---|---|---|---|---|
| w/o RGA | 86.31 | 85.95 | 81.18 | 76.56 |
| w/ RGA1 | 87.59 | 87.46 | 82.99 | 78.86 |
| w/ RGA2 | 87.32 | 87.12 | 82.61 | 78.32 |
| STF-HFNet | 88.34 | 88.27 | 84.02 | 80.14 |

### C. Internal Structure and Visualisation Analysis of HCF Module

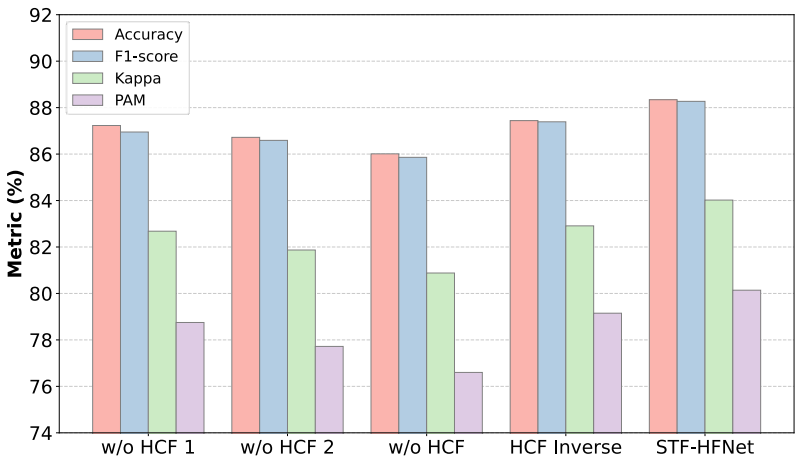

Figure 13: Ablation Study on HCF Module (Where HCF1 and HCF2 denote HCF variants only using channel and spatial attention, respectively.).

HCF improves model performance by acting as a cross-modal gating module that suppresses modality-specific noise and emphasises task-relevant joint features, as shown in Fig. 13. Removing either component (w/o HCF 1 or w/o HCF 2) reduces accuracy, and thus demonstrates that both channel-wise semantic filtering and temporal saliency localisation are indispensable. The more significant performance drop in the w/o HCF 2 setting further indicates that temporal saliency is more critical for handling asynchronous physiological rhythms. The HCF inverse variant also confirms the importance of correct inductive bias matters: EEG provides stronger channel-wise semantic guidance, while cardiovascular signals offer a more reliable structural prior for temporal masking. Overall, the w/o HCF setting achieves the worst performance, which suggests that linear fusion fails to capture the nonlinear cross-modal coupling that HCF explicitly models.

Based on the feature visualisation results in Fig. 14, within the HCF module, we regard peripheral signals as spatially salient features to calibrate the most prominent key segments of emotional responses (e.g., heart rate mutations). Meanwhile, we treat EEG signals as channel-level semantic features: They contain information from multiple frequency bands and spatial sources, which the network encodes into feature maps from multiple angles, corresponding to discriminative clues for different frequency bands and spatial patterns. Building on this, we leverage the spatial saliency of peripheral signals to weight and fuse the semantics of EEG channels, thereby constructing a gating mechanism. This mechanism can suppress background noise and irrelevant rhythms during non-response periods, highlight neural representations synchronised with peripheral arousal, and further enhance the consistency of cross-modal alignment and the ability of fusion expression.

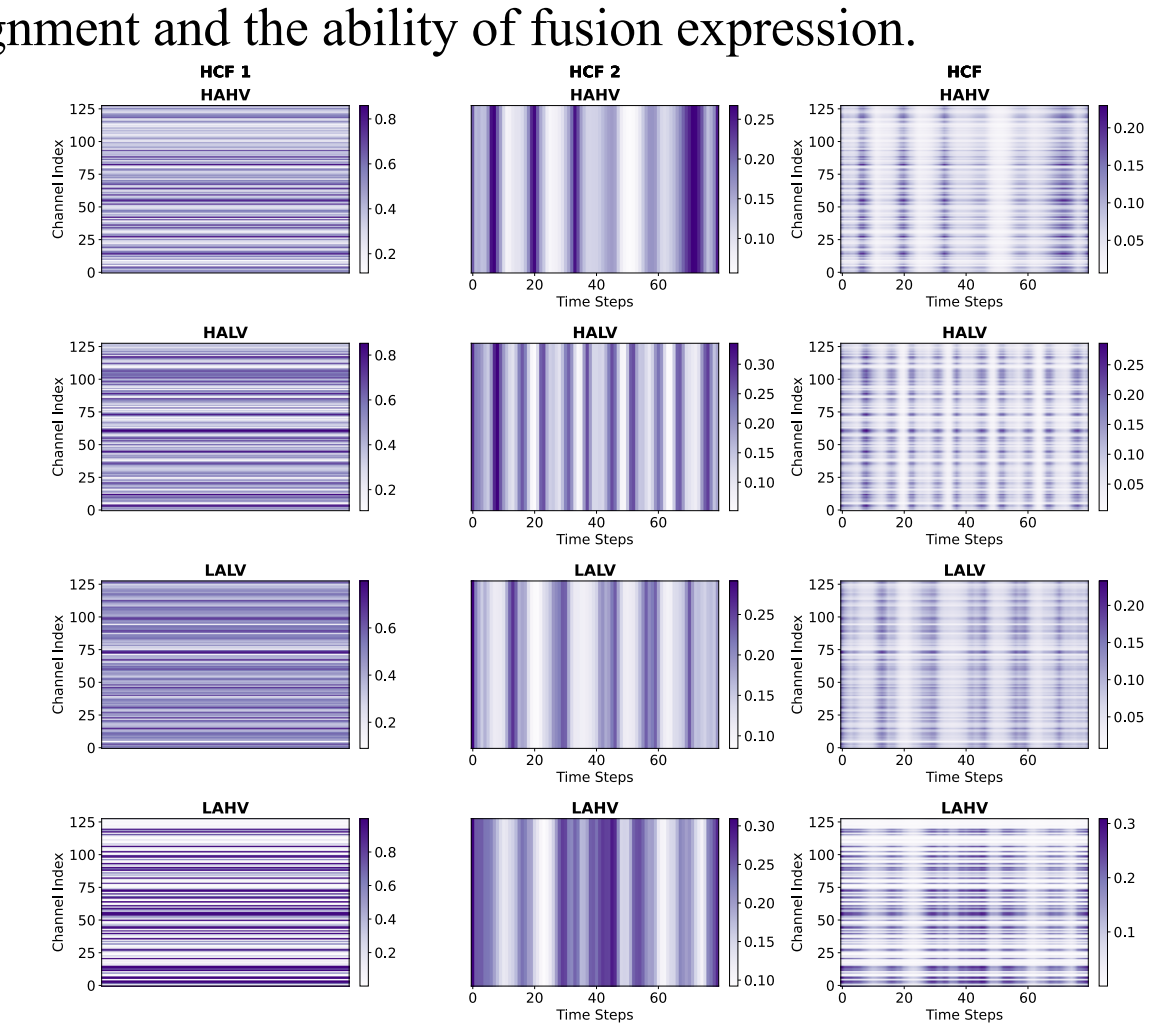

Figure 14: Visual Analysis of Attention Weights of the HCF Module.

## VII. CONCLUSION

This paper establishes the feasibility of wearable olfactory affective computing for continuous emotion monitoring. The main findings are as follows. We verify that olfactory stimuli can be effectively mapped to the arousal-valence space, and propose a three-classification task (HAHV, HALV, and MAMV) as a more rigorous benchmark for wearable sensing. We propose STF-HFNet, which models global frequency, and temporal and spatial dependencies while aligning brain and bodily representations to capture central-peripheral coupling during odor inhalation. This approach reflects the neurophysiological dynamics that underlie emotion recognition and enhances the accuracy of emotion decoding. Through EEG topographic visualisation and interpretability analysis, we confirm that the model captures biologically consistent markers. Overall, this work provides a non-intrusive, low-cognitive-load paradigm for wearable affective computing and, in turn, advances the field of continuous mental health monitoring.